\documentstyle[aps,preprint]{revtex}
\begin{document}
\draft
\tighten
\preprint{\vbox{
                \hfill TIT/HEP-395/NP 
 }}
\title{Properties of $N^*$(1535) at Finite Density in the Extended 
Parity-Doublet Models}
\author{Hungchong Kim \footnote{Email : hung@th.phys.titech.ac.jp},
D. Jido \footnote{Email : jido@th.phys.titech.ac.jp} and
M. Oka \footnote{Email : oka@th.phys.titech.ac.jp}}
\address{Department of Physics, Tokyo Institute of Technology, Meguro, 
         Tokyo 152-8551 Japan}

\maketitle
\begin{abstract}
We improve so called ``naive'' and ``mirror'' models for the 
positive and negative parity nucleons, $N$ and $N^{*}$,  
by introducing nonlinear terms allowed by chiral symmetry. 
Both models in this improvement reproduce the observed
nucleon axial charge in free space  and reveal interesting density 
dependence of the axial charges for $N$ and $N^*$, and 
the doublet masses. 
A remarkable difference between the two models is 
found in the off-diagonal axial charge, $g_{ANN^*}$,
which could appear either as suppression or as enhancement of $N^* \rightarrow
\pi N$ decay in the medium. 
\end{abstract}

\section{INTRODUCTION}
\label{sec:intro}

The negative-parity nucleon resonance, $N^*$(1535) , is
believed to play a dominant role in $\eta$ productions via
electromagnetic or hadronic probes~\cite{expts}.
There have been some attempts to understand 
$N^*$(1535)  resonance as well as its decaying properties 
based on effective models~\cite{eta_the}.  More recently, the resonance
has been studied in the framework of QCD sum rules by using 
the interpolating
field either with a covariant derivative~\cite{kim} or 
without~\cite{oldjido}.   

One interesting attempt in the study of $N^*$  is
to construct a chiral model where the resonance is
regarded as a chiral partner of a nucleon.
As chiral symmetry and its spontaneous breaking is important in
describing low energy nuclear physics, it seems natural to extend the
baryonic sector of chiral Lagrangian and include the
negative-parity baryon. 

Indeed, such extension was proposed by DeTar and 
Kunihiro~\cite{detar}.  In their model, the negative-parity baryon, 
under the chiral rotation,  is allowed to transform in the opposite
direction of the way that the positive-parity baryon transforms.
In this assignment of the chiral transformation which we call 
``mirror assignment'',  the nucleon 
may have a nonzero mass, $m_0$, even when the chiral symmetry is restored.
This seems to support the results from lattice calculations at finite
temperature~\cite{temp}.  One interesting prediction of this
model is that the axial charge of
$N^*$ has the opposite sign of the nucleon axial charge, which seems
to agree with the QCD sum rule approach proposed by 
Lee and Kim~\cite{kim}.

On the other hand, as pointed out by Jido {\it et al.}~\cite{jido}, one can 
also construct a chiral model by  allowing the negative-parity
baryon to transform in the same way as the positive-parity baryon under
the chiral rotation.  Under this assignment which we call ``naive 
assignment'', the $\pi N N^*$ coupling is zero and the $N^*$ axial
charge has the same sign as the nucleon. The vanishing
$\pi NN^*$ coupling is at least qualitatively consistent with the
suppression of the coupling observed in experiment. This way of 
including
the negative-parity baryon  leads to two independent 
linear sigma models and  the masses of $N$ and $N^*$ vanish
as the chiral symmetry is restored. This model 
yields the results consistent with the QCD sum rule approach proposed by
Jido {\it et al.}~\cite{oldjido}.

At present, a question remains as to which assignment of chiral symmetry is 
realized in the real world. Finding the answer
will help us not only to understand
$N^*$ itself but also to make reliable predictions on
$N^*$ properties as chiral symmetry is restored.  
Certainly the relative sign of 
the $N^*$ axial charge, if measured, can be used to determine the
realistic model.  But it will be difficult to measure the
axial charge because $N^*$ decays strongly to $\eta N$ or $\pi N$.    
Instead, it will be useful to pursue other perspectives of the
two models. 

To do so, we propose in this work a minimal extension of each model by 
introducing additional terms in the Lagrangian allowed by respective 
chiral symmetry.
Similar extension of the original linear sigma model,
normally known as modified linear sigma model,
has been proposed in order to have the axial charge of nucleon
greater than unity~\cite{akh}.  Such extension will be interesting in our
nucleon doublet models as it may provide different predictions
for the doublet masses and their couplings to pion as
the chiral symmetry is restored, which then may be
measured in pion photoproductions in nuclei or possibly in heavy-ion
collisions.   

Our paper is organized as follows.  In Section~\ref{sec:naive}, we 
propose an extension of the naive model after briefly reviewing 
the original naive model.  We discuss why the original naive 
model needs to be improved and make a nonlinear transformation of the 
extended model to describe $\pi$ interactions in the pseudovector 
coupling scheme.  In Section~\ref{sec:mirror}, we provide a similar 
extension for the mirror model.  
In Section~\ref{sec:res}, we apply our models to a system with
finite baryon density and see how the doublet masses and the axial
charges are changed 
as respective chiral symmetry is being restored. 
Throughout this paper, $N^{*}$ denotes $N(1535)$.

\section{Extension of the Naive Model}
\label{sec:naive}

The well-known linear sigma model of Gell-Mann and L\'evy can be 
constructed by requiring its Lagrangian to be invariant under the
following chiral transformations,
\begin{eqnarray}
N_1 \rightarrow N_1  - i{\mbox{\boldmath $\alpha$}} \cdot 
 {{\mbox{\boldmath $\tau$}} \over 2}
\gamma_5 N_1\;; \quad
{\mbox {\boldmath $\pi$}} \rightarrow   
{\mbox {\boldmath $\pi$}} +   
{\mbox {\boldmath $\alpha$}} \sigma \;; \quad 
\sigma \rightarrow \sigma - {\mbox {\boldmath $\pi$}} \cdot 
{\mbox {\boldmath $\alpha$}}\ .
\label{naivef1}
\end{eqnarray}
Here {\mbox {\boldmath $\alpha$}} denotes  an infinitesimal vector for the 
transformations and {\mbox{\boldmath $\tau$}} refers to the Pauli matrices
acting on the isospin space. Of course, one has to impose
the symmetry under the vector transformation in constructing
the Lagrangian.
One possible way to include the negative-parity baryon, $N_2$,
in the linear sigma model
is to construct a Lagrangian invariant under the further transformation of
\begin{eqnarray}
N_2 \rightarrow N_2  - i{\mbox{\boldmath $\alpha$}} \cdot 
 {{\mbox{\boldmath $\tau$}} \over 2}
\gamma_5 N_2 \ . 
\label{naivef2}
\end{eqnarray}
One can construct a simple Lagrangian
invariant under these chiral transformations~\cite{jido}, 

\begin{eqnarray}
{\cal L}_N^{0} =  &&\sum_{j=1,2} \left [{\bar N_j} i \not\!\partial N_j -
             a_j{\bar N_j} (\sigma + i \gamma_5 {\mbox {\boldmath $\tau$}}
             \cdot {\mbox {\boldmath $\pi$}}) N_j \right] \nonumber \  \\ 
            &&-a_3\left[{\bar N_1} (\gamma_5\sigma + i {\mbox {\boldmath $\tau$}}
             \cdot {\mbox {\boldmath $\pi$}}) N_2  -
             {\bar N_2} (\gamma_5\sigma + i {\mbox {\boldmath $\tau$}}
             \cdot {\mbox {\boldmath $\pi$}}) N_1 \right ] + {\cal L}_m\ .
\label{onav}
\end{eqnarray}  
Here we do not specify the mesonic part ${\cal L}_m$ 
because it is not needed in  most discussions below. 
This model is called as ``naive'' because the Lagrangian
is constructed by the naive extension of chiral symmetry to the
negative-parity baryon given in Eq.~(\ref{naivef2}).

Normally, a nucleon acquires its mass by spontaneous breaking
of chiral symmetry. 
In this case with the additional baryonic field, 
we have a mass matrix with nonzero off-diagonal elements (namely the 
term proportional to $a_3$) in the baryon doublet basis ${ N_1 \choose 
N_2 } $.  The physical masses for $N$ and $N^*$ can be obtained by 
diagonalizing the mass matrix.  However, under the diagonalization, 
the matrix representing the interactions between pion and baryon 
doublets is also diagonalized, which therefore leads to the coupling, 
$g_{\pi N N^*} = 0$.  This implies that this model is nothing but the 
sum of two independent sigma models.  Furthermore, 
using the physical values of the doublet masses 
and the pion decay constant $f_{\pi}$, we find $g_{\pi NN}=10$, which 
is about 30 \% smaller than what is
normally known.  This discrepancy is because the axial charges 
for both $N$ and $N^{*}$ are unity  independent  of the parameters
$a_{i}$ or $\langle \sigma \rangle$. Since, within this model, no other 
adjustments are allowed 
to achieve consistency with phenomenology in the tree level,  
this naive model needs to be improved.
 
The meson-baryon couplings of the Lagrangian Eq.~(\ref{onav}) are
linear in the meson field. 
One natural extension of the naive model is to add 
quadratic meson-baryon interactions to ${\cal L}_N^0$, 
\begin{eqnarray}
{\cal L}_N^1 =&& \sum_{j=1,2} b_j {\bar N}_j  \left [ ( 
         {{\mbox{\boldmath $\tau$}} \over 2} \cdot {\mbox {\boldmath $\pi$}} 
        \not\!\partial \sigma  - \sigma{{\mbox{\boldmath $\tau$}} \over 2}
        \cdot \not\!\partial {\mbox {\boldmath $\pi$}} ) \gamma_5
        - {{\mbox{\boldmath $\tau$}} \over 2}\cdot ({\mbox {\boldmath $\pi$}}
        \times \not\!\partial {\mbox {\boldmath $\pi$}}) \right ] N_j
\nonumber \ \\
          && + b_3 \left \{ {\bar N}_1  \left [ (
         {{\mbox{\boldmath $\tau$}} \over 2} \cdot {\mbox {\boldmath $\pi$}}
        \not\!\partial \sigma  - \sigma{{\mbox{\boldmath $\tau$}} \over 2}
        \cdot \not\!\partial {\mbox {\boldmath $\pi$}} ) 
        - {{\mbox{\boldmath $\tau$}} \over 2} \cdot ({\mbox {\boldmath $\pi$}}
        \times \not\!\partial {\mbox {\boldmath $\pi$}}) \gamma_5 \right ] N_2 
        + h.c. \right \}\label{n2}\ .
\end{eqnarray}
It is easy to see that these terms, whose form being
vector$-$vector plus axial-vector$-$axial-vector,
are invariant under the chiral transformations, 
Eqs.~(\ref{naivef1}) and (\ref{naivef2}).   
Note that these are constructed by using the least number of 
meson fields and derivatives among all possibilities
allowed by the naive assignment of chiral symmetry.  
Chiral symmetry does not prevent from having other
terms containing more meson fields or more derivatives. 
In this sense, our extension should be
regarded as  minimal. 
The truncation in number of mesons and derivatives makes sense
when  contributions from the additional terms are not so large.  
Similar modification of the linear sigma model has been proposed
by Lee~\cite{lee} and later used  
by Akhmedov~\cite{akh} in an effort to have the
axial charge greater than 1 in the tree level.  
In our case, because of  the negative-parity baryon,
we have the off-diagonal terms in addition. 
As the nucleon mass matrix is unchanged by including ${\cal L}_N^1$,
the previous 
mass eigenstates  still form the physical basis in this extension. 

However, in this extension, the pion couples to the nucleons
both in  pseudoscalar type and in pseudovector type.   
It is well known~\cite{brian} that the pseudoscalar coupling scheme 
leads to too large $s$-wave $\pi N$  scattering length from
the Born terms, which then needs to be 
reduced by the $t$-channel sigma exchange between the pion and the
nucleon. 
This sensitive cancellation however is not necessary
when one uses the pseudovector coupling scheme obtained from nonlinear
transformations of the linear sigma model. Furthermore, a Hartree-Fock
calculation of nuclear matter shows~\cite{horowitz} that the pseudoscalar
coupling scheme is problematic in describing normal nuclear matter. 
Therefore, it is favorable to have a model in the pseudovector coupling scheme
only.  This can be achieved by the nonlinear realization of chiral
symmetry, as originally studied by Weinberg~\cite{wein}.

Motivated by Weinberg, we take the following nonlinear transformations,
\begin{eqnarray}
\psi_j &=&{ 1 \over \sqrt{ 1 +  \left ({\pi^\prime \over 2 f_\pi}\right )^2}}
        \left [ 1 -{i \over 2 f_\pi} {\mbox {\boldmath $\pi^\prime$}} \cdot 
         {\mbox {\boldmath $\tau$}} \gamma_5 \right ] N_j ~~{\rm where}~~j=1,2
\nonumber\ ,\\
{{\mbox {\boldmath $\pi^\prime$}}\over 2f_\pi} &=& 
{{\mbox {\boldmath $\pi$}} \over \sqrt{\sigma^2 + \pi^2} + \sigma }
\;;\quad
\sigma^\prime =\sqrt{\sigma^2 + \pi^2}\label{nonlinear}\ .
\end{eqnarray}
We want to stress that the negative-parity baryon is
transformed similarly to the positive-parity baryon so that
the naive assignment of chiral symmetry is realized nonlinearly. 
After applying the nonlinear transformations to the Lagrangian 
(${\cal L}_N^0 + {\cal L}_N^1$) and implementing the spontaneous symmetry 
breaking, we obtain 
\begin{eqnarray}
{\cal L}^{NL}_{N} &=& {\bar \psi} i \not \! \partial \psi +
{ 1 \over  1 +  \left ({\pi^\prime \over 2 f_\pi}\right )^2}
{\bar \psi} \not \! \partial {\mbox {\boldmath $\pi^\prime$}}\cdot
{\mbox {\boldmath $\tau$}} \gamma _5 {\bf \Pi} \psi
-{ (1 / 2f_\pi)^2  \over  1 +  
\left ({\pi^\prime \over 2 f_\pi}\right )^2 } 
{\bar \psi}
{\mbox {\boldmath $\tau$}} \cdot ({\mbox {\boldmath $\pi^\prime$}}\times
\not\!\partial {\mbox {\boldmath $\pi^\prime$ }}) \psi\nonumber\ \\
&&-{\bar \psi} {\bf M} \psi + {\cal L}_m\label{nlag}\ ,
\end{eqnarray} 
where we have introduced the column vector, 
$\psi \equiv  { \psi_1 \choose \psi_2 } $.  In this
basis, the pion
coupling and the baryon mass matrices are defined respectively as
\begin{eqnarray}
{\bf \Pi} =  {1 \over 2 f_\pi} \pmatrix 
           { 1-b_1 \sigma_0^2 & -b_3 \sigma_0^2 \gamma _5 \cr
           -b_3 \sigma_0^2 \gamma_5  & 1 - b_2 \sigma_0^2 \cr }
\;; \quad
{\bf M} = \sigma_0 \pmatrix { a_1 & a_3 \gamma_5 \cr
                                   -a_3 \gamma_5 & a_2  \cr } \ ,
\end{eqnarray}
where $\sigma_0$ denotes the expectation value of $\sigma^\prime$
whose value in vacuum is the pion decay constant $f_\pi$, and 
the fluctuation of $\sigma^\prime$
field around $\sigma_0$ is neglected. 
Note that the nonlinear transformations eliminate the pseudoscalar 
coupling of the pion. The $\sigma_0$
dependence of the couplings is now explicit, while 
nonlinear sigma models constructed without facilitating
the linear sigma model lead to  no explicit $\sigma_0$ 
dependence of the couplings and therefore we may not be able to study
phenomena related to  chiral symmetry restoration.

Now, because of the new terms involving two 
mesons, the pion-nucleon coupling matrix is different from the mass 
matrix.  Diagonalization of the mass matrix does not simultaneously 
diagonalize the coupling matrix.  Therefore, the $\pi N N^*$ coupling 
picks up a nonzero value in the tree level, which is certainly 
consistent with experimental facts.  Moreover $\sigma_0$ is 
just an overall factor in the mass matrix.  Thus the unitary matrix 
which diagonalizes the mass matrix does not depend on $\sigma_0$.  
It implies that, after the diagonalization, the coupling matrix will 
take the form,
\begin{eqnarray}
{\bf \Pi} \rightarrow  {1 \over 2 f_\pi} \pmatrix 
           { 1+B_1 \sigma_0^2 & B_2 \sigma_0^2 \gamma _5 \cr
           B_2 \sigma_0^2 \gamma_5  & 1 + B_3 \sigma_0^2 \cr }
\label{dpi}\ 
\end{eqnarray}
with some constants, $B_i$ ($i=1,2,3$).
The matrix for the axial charges can be simply read off, giving
\begin{eqnarray}
   \pmatrix
           { g_{ANN} & g_{ANN^*} \cr
           g_{ANN^*}  & g_{AN^*N^*} \cr }
\equiv
          \pmatrix
           { 1+B_1 \sigma_0^2 & B_2 \sigma_0^2  \cr
           B_2 \sigma_0^2  & 1 + B_3 \sigma_0^2 \cr }\ .
\end{eqnarray}

The constants, $B_1$ and $B_2$, can be determined by fitting experimental values
for $\pi NN$ and $\pi N N^*$ couplings respectively.
Specifically, the partial decay width
for $N^* \rightarrow \pi N$ 
can be evaluated straightforwardly from the $\pi N$ self-energy of $N^*$,
\begin{eqnarray}
\Gamma_{\pi N} = {3 f_{\pi N N^*}^2 \over 8 \pi M_{-}^2 } |{\bf k}|
(M_{-}-M_+) \left [ (M_{-}+M_{+})(M_{-}^2-M_{+}^2-m_\pi^2) + 
2 M_{+} m_\pi^2 \right ]
\label{coup}
\end{eqnarray}
where $|{\bf k}|$ is the momentum carried by the emitted pion and
$f_{\pi NN^*} \equiv B_2 \sigma_0^2 /2f_\pi$ is the off-diagonal pseudovector
coupling.
From $\Gamma^{exp}_{\pi N} \sim 70$ MeV,
we obtain 
$f_{\pi N N^*} = 1.17$ GeV$^{-1}$, which then yields $B_2=25.6$ GeV$^{-2}$
when the vacuum expectation value, $\sigma_0 
= 93$ MeV, is used.
Also from the diagonal pseudovector coupling 
$f_{\pi N N} \equiv (1+B_1 \sigma_0^2)/2f_\pi 
= 6.77$ GeV$^{-1}$, we find $B_1= 30.06$ GeV$^{-2}$.
Note that the small deviation of $g_{ANN}$ from unity and the suppressed
off-diagonal coupling $f_{\pi N N^*}$ make $B_1 \sigma_0^2 = 0.26$ and
$B_2 \sigma_0^2=0.217$, much smaller than unity, justifying the truncation
in number of meson fields in Eqs.~(\ref{onav}) and (\ref{n2}).

The masses of $N$ and $N^*$ are given as
\begin{equation}
M_+ = A_1 \sigma_0 \;; \quad M_- = A_2 \sigma_0\ ,
\end{equation}
where the constants, $A_1$ and $A_2$, are determined from 
the physical masses.
When $\sigma_0$ is changed
in the medium, $M_{\pm}$ will be changed linearly with $\sigma_0$.
A similar scaling is proposed by 
Brown and Rho~\cite{brown}
where all hadron masses except for the  Goldstone bosons are claimed 
to be scaled with the pion decay constant in the nuclear medium.

\section{Extension of the Mirror Model}
\label{sec:mirror}

Another way of constructing the nucleon parity doublet model is to introduce 
a Lagrangian invariant under the ``mirror assignment'' of chiral
symmetry~\cite{mirr}, originally proposed by Lee~\cite{lee} and
later developed by  DeTar and Kunihiro(DK)~\cite{detar}.
Under the mirror assignment, the negative-parity baryon is allowed
to transform in the opposite direction to the way that the
positive-parity baryon is transformed, namely 
\begin{eqnarray}
N_1 &\rightarrow& N_1  - i{\mbox{\boldmath $\alpha$}} \cdot 
 {{\mbox{\boldmath $\tau$}} \over 2}
\gamma_5 N_1 \;; \quad 
N_2 \rightarrow N_2  + i{\mbox{\boldmath $\alpha$}} \cdot 
 {{\mbox{\boldmath $\tau$}} \over 2}
\gamma_5 N_2 \nonumber\ ,\\
{\mbox {\boldmath $\pi$}} &\rightarrow&   
{\mbox {\boldmath $\pi$}} +   
{\mbox {\boldmath $\alpha$}} \sigma \;; \quad 
\sigma \rightarrow \sigma - {\mbox {\boldmath $\pi$}} \cdot 
{\mbox {\boldmath $\alpha$}}\ .
\label{mirrorf}
\end{eqnarray}
Lagrangian invariant under this transformation is
\begin{eqnarray}
{\cal L}_M^0 &=& {\bar N}_d i \not\! \partial N_d - g_1 {\bar N}_d 
(\sigma + 
i {\mbox{\boldmath $\pi$}} \cdot {\mbox{\boldmath $\tau$}} \gamma_5 \rho_3 )
N_d 
+
g_2 {\bar N}_d
(\sigma \rho_3+ 
i {\mbox{\boldmath $\pi$}} \cdot {\mbox{\boldmath $\tau$}} \gamma_5 )
N_d\nonumber\ \\
&&- i m_0 {\bar N}_d \rho_2 \gamma_5 N_d 
+{\cal L}_m \label{ml}
\end{eqnarray}
where we have introduced 
the parity doublet 
$N_d \equiv  { N_1 \choose N_2 } $ for notational simplicity, and
the Pauli matrices $\rho_i~(i=1,2,3)$ which
act on the nucleon parity doublets.   Note that because of the 
mirror assignment, the Lagrangian is allowed to have the term containing
$m_0$ which constitutes the off-diagonal element of the
mass matrix.

In this case, unlike the naive case, the mass and the coupling 
matrices are not
simultaneously diagonalized and we can have a nonzero value for
$g_{\pi N N^*}$ without further introducing nonlinear terms.   
Using the experimental
 value for $g_{\pi N N^*}$, $M_+$ and
$M_-$, DK found,
\begin{eqnarray}
g_1 = 13\;; \quad  g_2 = 3.2 \;; \quad m_0 = 0.27~{\rm GeV} \ .
\end{eqnarray}
This model is interesting because, in the restored phase of chiral
symmetry, the nucleons still have the nonzero mass, $m_0$.
The nonzero mass in the restored phase, though it is small, 
seems to be  
consistent with lattice calculations at finite 
temperature~\cite{temp}. In addition, this approach reveals interesting 
behavior of $g_{ANN}$ as a function of $\sigma_0$ .
However, this model predicts  $g_{\pi NN} = 9.8$.  
The well-known value of $g_{\pi NN}$ can not be incorporated
because the predicted axial charges for $N$ and $N^{*}$ are always 
less than 1 at the tree level.
Therefore, the mirror model in the
current form is somewhat too restrictive and needs to be improved.

One way to improve the mirror model is to take similar steps as in the
previous section.   
Under the mirror assignment of chiral symmetry,  
the simplest extension of the model is to include the followings,
\begin{eqnarray}
{\cal L}_M^1 =
&& g_3 {\bar N}_1 ( i \not \! \partial\sigma \gamma_5 + 
{\mbox{\boldmath $\tau$}} \cdot \not \! \partial {\mbox{\boldmath $\pi$}})
N_2 + h.c. \nonumber\ \\
&& + g_4 {\bar N}_1  \left [ ( 
         {{\mbox{\boldmath $\tau$}} \over 2} \cdot {\mbox {\boldmath $\pi$}} 
        \not\!\partial \sigma  - \sigma{{\mbox{\boldmath $\tau$}} \over 2}
        \cdot \not\!\partial {\mbox {\boldmath $\pi$}} ) \gamma_5
        - {{\mbox{\boldmath $\tau$}} \over 2}\cdot ({\mbox {\boldmath $\pi$}}
        \times \not\!\partial {\mbox {\boldmath $\pi$}}) \right ] N_1
\nonumber \ \\
&& + g_5 {\bar N}_2  \left [ -( 
         {{\mbox{\boldmath $\tau$}} \over 2} \cdot {\mbox {\boldmath $\pi$}} 
        \not\!\partial \sigma  - \sigma{{\mbox{\boldmath $\tau$}} \over 2}
        \cdot \not\!\partial {\mbox {\boldmath $\pi$}} ) \gamma_5
        - {{\mbox{\boldmath $\tau$}} \over 2}\cdot ({\mbox {\boldmath $\pi$}}
        \times \not\!\partial {\mbox {\boldmath $\pi$}}) \right ] N_2
\label{m2}\ .
\end{eqnarray}
Here the second and third lines contain two mesons similarly to Eq.~(\ref{n2}).
Note the sign of the axial-vector coupling term in the third line is 
opposite to 
that in the second line.  This is because of the mirror assignment of
chiral symmetry for the negative-parity baryon.   The terms containing
one meson are new and these are not allowed in the naive assignment.
Of course, under chiral symmetry, terms with more mesons or more derivatives
are also allowed.  Therefore, this extension as in the naive case should
be regarded as minimal. As before, the truncation in number of mesons and
derivatives can be justified if the contributions of these corrections 
are not so large. 

Again, in this extension, the pion couples to the nucleons
both in pseudoscalar type and in pseudovector type. 
Nonlinear transformations are required to eliminate the pseudoscalar
coupling in favor of the pseudovector coupling. 
The mirror assignment of 
chiral symmetry is realized with slightly different nonlinear transformation 
for the parity doublets, 
\begin{eqnarray}
\psi &=&{ 1 \over \sqrt{ 1 +  \left ({\pi^\prime \over 2 f_\pi}\right )^2}}
        \left [ 1 -\rho_3 {i \over 2 f_\pi} {\mbox {\boldmath $\pi^\prime$}} 
\cdot 
         {\mbox {\boldmath $\tau$}} \gamma_5 \right ] N_d \ ,
\end{eqnarray} 
while $\pi$ and $\sigma$ are transformed as before. 
Note that $\rho_3$ indicates the sign difference for 
the negative-parity baryon. 
The nonlinear transformations and subsequent spontaneous
symmetry breaking  takes the Lagrangian (${\cal L}^0_M + {\cal L}^1_M$) 
into the same form as 
in Eq.~(\ref{nlag})
but the pion coupling and mass matrices now take different forms,
\begin{eqnarray}
{\bf \Pi} =  {1 \over 2 f_\pi} \pmatrix 
           { 1 - g_4 \sigma_0^2 & 2 g_3 \sigma_0 \gamma _5 \cr
           2 g_3 \sigma_0 \gamma_5  & -1- g_5 \sigma_0^2  \cr }
\;; \quad
{\bf M} =  \pmatrix { (g_1 - g_2) \sigma_0  & m_0 \gamma_5 \cr
                             -m_0 \gamma_5 & (g_1 + g_2)\sigma_0 \cr } \ .
\end{eqnarray}
Since there is no dependence on $g_3$, $g_4$ and $g_5$ in the mass
matrix, the mass eigenstates are the same as before the extension.
The mass matrix is diagonalized by the unitary matrix
\begin{eqnarray}
{\bf U} = {1 \over 2 {\rm cosh} \delta }
          \pmatrix { e^{{\delta \over 2}}  & e^{-{\delta \over 2}} \gamma_5 \cr
               e^{-{\delta \over 2}} \gamma_5  & -e^{{\delta \over 2}}  \cr
}~~ {\rm where}~~{\rm sinh}\delta = {g_1 \sigma_0 \over m_0}\ .
\end{eqnarray}
Note that the mixing 
angle $\delta$ depends on $\sigma_0$. This is in contrast
with the naive case.
The physical masses are obtained as 
\begin{eqnarray}
M_{\pm} = {\mp} g_2 \sigma_0 +
\sqrt{(g_1 \sigma_0)^2 + m_0^2 }\ .
\end{eqnarray}

To get the physical pion couplings, we apply the unitary transformation
also to ${\bf \Pi}$, which then leads to the axial charges,
\begin{eqnarray}
g_{ANN}&=&{\rm tanh}\delta
- { g_4 {\sigma_0}^2 e^\delta - g_5 {\sigma_0}^2
e^{-\delta} \over 2 {\rm cosh}\delta }
+ {2 \sigma_0 g_3 \over {\rm cosh}\delta }\nonumber\ ,\\
g_{ANN^*}&=&{2-{\sigma_0}^2 (g_4+g_5) \over 2{\rm cosh}\delta}
-2 \sigma_0 g_3 {\rm tanh}\delta\nonumber\ ,\\
g_{AN^*N^*}&=&-{\rm tanh}\delta
- { g_4 {\sigma_0}^2 e^{-\delta} - g_5 {\sigma_0}^2
e^{\delta} \over 2 {\rm cosh}\delta }
- {2 \sigma_0 g_3 \over {\rm cosh}\delta}\label{mcoup}\ .
\end{eqnarray}
Unfortunately, we have six undetermined parameters in this extension while
there are only four experimental inputs, 
\begin{eqnarray}
&&M_- = 1.535~{\rm GeV} \;; \quad M_+ = 0.939~{\rm GeV} \nonumber\ ,\\
&&g_{A NN} = 1.26 \;; 
\quad g_{A NN^*} = 0.217 \ .
\end{eqnarray}
One more constraint can be imposed by noting that $g_{ANN}$ and $g_{AN^*N^*}$
are differed only by the quadratic terms in meson fields.
Since the quadratic and linear terms in meson fields were introduced
in a way to improve the linear sigma model, the difference
should be small.  Otherwise, the truncation in number of
meson fields can not be justified.  This means that 
$g_{ANN}$ can be assumed to be similar or even equal in magnitude with
$g_{AN^*N^*}$. Also, as in the original DK model, we 
expect the sign of
$g_{AN^*N^*}$ to be the opposite of $g_{ANN}$.
In Section~\ref{sec:res}, we will present the results with
the condition,  
$g_{ANN} = -g_{AN^*N^*}$, which yields $g_4 = g_5$. 
The reliability of this constraint can be checked by allowing a small
deviation $\epsilon$ such that
\begin{equation}
g_{ANN} = -g_{AN^*N^*} + \epsilon\label{eps}\ .
\end{equation}

One more freedom in our model will be fixed by
taking $m_0$ as an adjustable parameter 
for our predictions.
In principle, $m_0$
should be determined for example by lattice calculations.
In the DK model,  $g_{ANN}$ 
is proportional only to 
${\rm tanh}\delta$.  Since their $m_0$ is fixed to a small value, 
$g_{ANN}$ is a very smooth function of  
$\sigma_0$. Practically one can start to see
a noticeable reduction of $g_{ANN}$ only when    
$\sigma_0$ is as small as $20$ MeV.
Therefore, the so called ``$g_{ANN}$ quenching'' at the normal nuclear
density can not be observed within their model.
However,  in our extension, $m_0$ is not necessarily fixed to a small
value and  it is possible that part of the $g_{ANN}$ quenching
can be driven by this mechanism.

\section{Results and Discussions}
\label{sec:res}
In the last two sections, we have improved   
the ``naive'' and ``mirror'' models 
for the nucleon parity 
doublets.  The additional chiral symmetric terms are introduced so that
the naive and mirror models can be consistent with phenomenology
in free space. One  important consequence  of those terms is
that physical quantities such as masses or couplings
are functions of 
$ \sigma_0$.    
As $\sigma_0$ is expected to be changed in the nuclear 
medium, our
formalism could reveal interesting density dependence of 
those quantities.

To make realistic predictions, we calculate 
$\sigma_0$ in Hartree approximation as
a function of the nuclear density.  A similar calculation for the
mirror model has been performed by Hatsuda and Prakash~\cite{hatsuda}.
Since we use the pseudovector
coupling for pion, Fock exchange terms are 
not expected to to be large~\cite{horowitz}. 
As the nucleus in the ground state does not contain $N^*$,
there is no Fermi energy for $N^*$ and 
the energy density in Hartree approximation takes
the form,
\begin{eqnarray}
{\cal E} = \lambda \left (  \sigma_0^2 -f_\pi^2 \right )^2 +
{\cal E}^+_v  + {\cal E}^-_v 
+ 4 \int_0^{k_F} {d^3 k \over (2 \pi)^3} \sqrt{{\bf k}^2 + {M^*_+}^2}\ ,
\end{eqnarray}
where $M^*_+$ denotes the positive-parity nucleon mass in the medium.
The constant $\lambda$ in front of the meson energy
is related with the mass of the $\sigma$ meson
via $\lambda = m_\sigma^2 / 8 f_\pi^2$.  In our calculation, we take 
$m_\sigma = 600$ MeV as in Ref.~\cite{hatsuda}. 
This $m_\sigma$ is also supported by  recent analysis of $\pi\pi$-scattering
phase shift~\cite{ishida}. The factor 4 in the last term indicates
that we are considering the  symmetric nuclear matter. 
We do not include the $\omega$ meson term interacting with nucleons
simply because it does not participate in determining $\sigma_0$ in matter.
The vacuum contributions from the positive-(negative-)parity nucleon
to the energy density,  ${\cal E}^+_v$ (${\cal E}^-_v$), diverge.
Suitable counter terms~\cite{brian} are introduced to obtain 
\begin{eqnarray}
{\cal E}^{\pm}_v &=& -{1 \over 4 \pi^2} \Big [ {M^*_{\pm}}^4 {\rm ln} 
{M^*_{\pm} \over M_{\pm}} -{M_{\pm}}^3(M^*_{\pm} -M_{\pm}) 
- {7 \over 2}(M^*_{\pm} -M_{\pm})^2 {M_{\pm}}^2
\nonumber\ \\
&-& {13 \over 3} M_{\pm}(M^*_{\pm} -M_{\pm})^3 
-{25 \over 12} 
(M^*_{\pm} -M_{\pm})^4  \Big ]\ .
\end{eqnarray}
Now we minimize the energy density to obtain
$\sigma_0$ at a certain density which is then used to
determine the values of $M^*_{\pm}$ and the axial charges,
$g_{A NN}$ and $g_{A N N^*}$.  

Figure~\ref{f1}~(a) shows our results from Hartree calculation for
$\sigma_0$ as a function of density. We consider
up to twice of the normal nuclear density, a range where 
this effective model approach is reliable. 
The two models clearly predict $\sigma_0$ to
decrease as the density increases.  
The curve for the mirror case 
is obtained by using $m_0=0.5$ GeV.  Larger $m_0$ leads to
slow decreasing rate for $\sigma_0$ but  
dependence on $m_0$ is not strong. 
The slope of $\sigma_0$ at $\rho=0$ is special as it
is related to $\pi N$-sigma
term, $\Sigma_{\pi N}$, 
via $d\sigma_0^2 / d \rho |_{\rho =0}= -\Sigma_{\pi N} /m^2_\pi$.  
The two curves in Fig.~\ref{f1}~(a) yield $\Sigma_{\pi N} \sim 100$ MeV
($\sim 80$ MeV) 
for the naive case (the mirror case) which is larger than
the typical value $\Sigma_{\pi N} \sim 45$ MeV.   The slope at
the zero density however is somewhat sensitive to $m_0$. For $m_0 = 700$ MeV,
we find $\Sigma_{\pi N} \sim 60$ MeV which is close to 
the typical value.  In this regards, our model with larger $m_0$ seems to
be favored. Larger $m_0$ is also suggested by  
lattice calculations~\cite{temp}.

Using the density dependence of $\sigma_0$ as an 
input, we calculate the doublet masses which are shown in Fig.~\ref{f1}~(b).  
In the mirror case, we have used $m_0 = 0.5$ GeV.  
We have also tried other values of $m_0$ and found
that the changes are at most 10 MeV at $\rho_0$.
Also shown is the $\eta$-meson mass (dot-dashed line) in vacuum. 
The two models predict $M_{\pm}^*$ to decrease as the
density increases,  seemingly consistent with
the Brown-Rho scaling~\cite{brown}.  Since the decreasing rate for $M_+^*$
is different from that of $M_-^*$, we find that
$M_-^* -M_+^*$ is also getting smaller as the density increases,
indicating that $N^*$ cannot decay to $\eta N$ in the medium if
the mass of $\eta$ is assumed not to change in the medium.
This feature is  insensitive to $m_0$. 

It is interesting to note that our results support
the recent observation~\cite{newexp} where
$\eta$-productions from nuclei are found to be scaled with the nuclear
surface area.  In Ref.~\cite{newexp}, the scaling is 
explained by strong nuclear absorptions of $\eta$ produced inside
a nucleus. Here, our results provide an alternative explanation for
the scaling. 
According to our models,  $N^* \rightarrow \eta N$ decay is possible only
at low density region below  $\rho/\rho_0 \sim 0.3$. Therefore,
$\eta$ is produced only at the surface of the nucleus and thus
the experimentally  observed scaling can be explained.

We now turn to results for the axial charges. As discussed above,
we here present the results with the condition $g_{ANN}=-g_{AN^* N^*}$
and discuss the reliability.
In Figure~\ref{f2}~(a),  
the nucleon axial charge, $g_{ANN}$, is shown to decrease as $\rho$ increases.
The quenching in the naive case is basically what
Akhmedov found in his modified sigma model~\cite{akh}.
At $\rho = \rho_0$, $g_{ANN}$ is quenched by 8 \%.
For the mirror case, the quenching rate depends strongly on the choice
of $m_0$.  
The axial charge decreases more rapidly for larger $m_0$. 
This trend is basically unchanged when we allow  
nonzero deviation $\epsilon$ defined in Eq.~(\ref{eps}). 
For example, even with a very large value   $\epsilon  =\pm 0.5$,
$g_{ANN}$ is changed only by 2 \% at $\rho_0$.
Therefore, the mirror model provides another interesting mechanism for
the quenching of the axial charge in medium.
When $\sigma_0$ decreases towards the chiral restoration 
$\sigma_0 \rightarrow 0$,
$g_{ANN}$ approaches to 1 in
the naive case while
in the mirror case it approaches to zero, because these values are
determined by their chiral assignment.

More clear distinction between the two assignments can be seen from the 
off-diagonal axial charge, $g_{ANN^*}$.  As shown in 
Fig.~\ref{f2}~(b), the two models predict opposite trends for  
$g_{ANN^*}$ as the density increases.  Starting from 
0.217, $g_{ANN^*}$ in the naive case decreases while its value
in the mirror model  increases. The increasing rate in the mirror
case strongly depends on  $m_0$.  Small $m_0$ usually tends to flatten 
the curve.  Specifically, the result with $m_0 = 0.3$ GeV shows that 
$g_{ANN^*}$ is 0.294 at $\rho_0$.  Corresponding result from
the naive model is 0.15, about a factor of 2 smaller.
As is clear from Fig.~\ref{f2}~(b), the gap is much bigger for larger $m_0$.  
We have also checked the sensitivity to $\epsilon$. 
Depending on its sign, its effect appears with either rapid or slow
increasing rate. For example, $\epsilon = \pm 0.5$  
changes $g_{ANN^*}$ at $\rho_0$
by at most $\pm 20$ \% .
Therefore, the qualitative aspect of our result is unchanged by the
uncertainty of the model.
Unlike the diagonal axial 
charge $g_{ANN}$, the 
off-diagonal $g_{ANN^{*}}$ in the naive case goes to zero when the 
chiral symmetry is restored, while in the mirror case 
$g_{ANN^{*}}$ goes to 1.  This implies 
that $N^{*}$ is the chiral partner of $N$~\cite{jido} .  

As $g_{ANN^*}$ is  
related to the pseudovector coupling, $f_{\pi NN^*}$, via 
$2 f_\pi f_{\pi NN^*} = g_{ANN^*}$,  the two models 
predict either suppression or enhancement of $\pi N$ decay of $N^*$ in
the medium. Effects from the Pauli blocking are not expected to be large as
the emitting nucleon has momentum much larger than the Fermi momentum. 
Since the 
decay of $N^{*}$ to $\eta N$ will be strongly suppressed 
(recall the discussion above),
we expect 
that the main mesonic decay mode of $N^{*}$ in nuclei would 
be  $\pi N$. Whether that partial width increases or decreases could be
used in  
determining the realistic assignment of chiral symmetry.

To summarize, we have minimally extended the naive and mirror models 
for the nucleon parity-doublet by introducing nonlinear terms allowed 
by respective assignment of chiral symmetry.  A unique description of 
the pion coupling in terms of the pseudovector scheme has been achieved by 
the nonlinear transformations.  We have investigated how the two 
assignments of chiral symmetry can be differed in our extension by 
performing Hartree calculation of $\sigma_0$ in 
the nuclear matter.  Using the density dependence of $\sigma_0$, we
have calculated the doublet masses and the axial 
charges in the medium.  We have found that the mass difference, 
$M_-^*-M_+^*$, decreases as the density increases.  Therefore, $N^*$ 
cannot decay to $\eta N$ in the medium, agreeing with recent 
experimental observation~\cite{newexp}.  
Also both the models predict that the axial 
charge, $g_{ANN}$, is quenched in the medium.  
Most remarkable result in our work is that  the two models
are clearly distinguished by the 
density dependence of the off-diagonal axial charge, $g_{ANN^*}$.  
Therefore, the realistic assignment of chiral symmetry might be 
determined by studying $\pi N$ and $\eta N$ decays of $N^*$ in medium.

\acknowledgments
This work is supported in part by the Grant-in-Aid for scientific
research (C) (2) 08640356 and (A) (1) 08304024 
of  the Ministry of Education, Science, Sports and Culture of Japan.
The work of  H. Kim and D. Jido is supported by Research Fellowships of
the Japan Society for the Promotion of Science.

\begin{figure}
\caption{Figure (a) shows the density dependence of $\sigma_0$.  
The solid line is for the naive case and the dashed line
is for the mirror case.    Figure (b) shows our  predictions for $M_-^*$, 
$M_+^*$ and the mass difference $M_-^* - M_+^*$ as density increases.
The solid lines (dashed lines) are for the naive case (mirror case).
For the mirror case, $m_0=0.5$GeV is used.  The two curves
for $M_-^*-M_+^*$ are almost indistinguishable.
Also shown with the dot-dashed line is the mass of $\eta$ in free space.}
\label{f1}
\vspace{10pt}
\caption{Our prediction for axial charges.  The solid lines (dashed lines)
are the results from the naive case (mirror case).
Figure (a) is for the nucleon axial charge and (b) is for
the off-diagonal axial charge. The results from the mirror case
are shown  for $m_0 =$ 0.3,0.5 and 0.7 GeV  as indicated.}
\label{f2}
\end{figure} 
\eject

\setlength{\textwidth}{6.1in}   
\setlength{\textheight}{9.in}  
\begin{figure}
\centerline{%
\vbox to 2.4in{\vss
   \hbox to 3.3in{\includegraphics{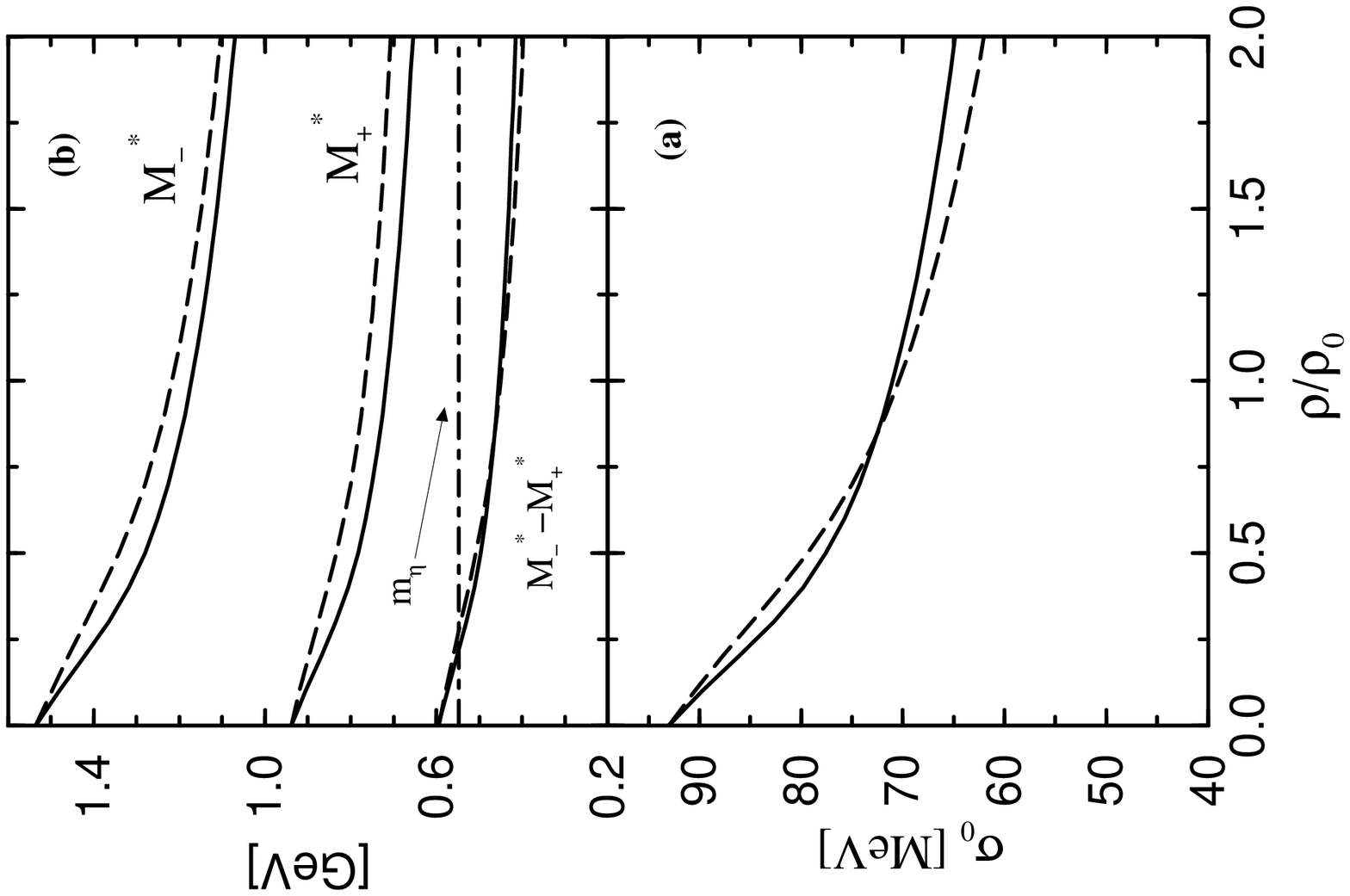}\hss}}
}
\bigskip
\vspace{400pt}
Figure 1
\end{figure}
\eject
\begin{figure}
\centerline{%
\vbox to 2.4in{\vss
   \hbox to 3.3in{\includegraphics{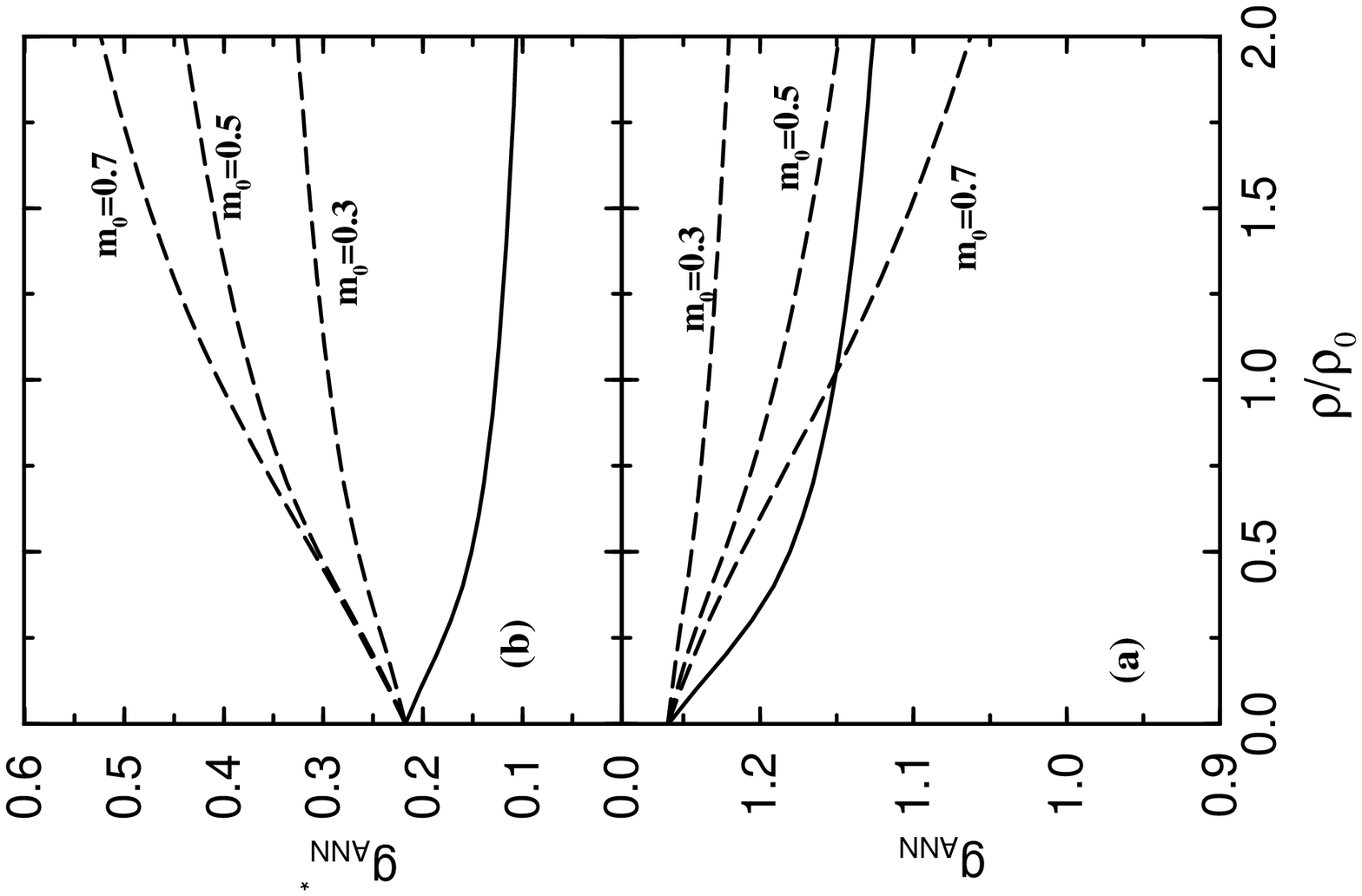}\hss}}
}
\bigskip
\vspace{400pt}
Figure 2
\end{figure}
\end{document}